\documentclass[sigconf, nonacm]{acmart}

\usepackage{tabularx}
\usepackage{tikz}
\usepackage[edges]{forest}
\usepackage{pgfplots}
\usepackage{graphicx} 
\pgfplotsset{compat=1.18}

\begin{document}

\title{Large Language Model Agent in Financial Trading: A Survey}

\author{Han Ding}
\authornote{All authors contributed equally to this research. Order is random.}
\email{hd2412@columbia.edu}
\affiliation{%
  \institution{Columbia University}
  \streetaddress{116th and Broadway}
  \city{New York}
  \state{NY}
  \country{USA}
  \postcode{10027}
}
\author{Yinheng Li}
\authornotemark[1]
\email{yl4039@columbia.edu}
\affiliation{%
  \institution{Columbia University}
  \streetaddress{116th and Broadway}
  \city{New York}
  \state{NY}
  \country{USA}
  \postcode{10027}
}

\author{Junhao Wang}
\authornotemark[1]
\email{jw3668@columbia.edu}
\affiliation{%
  \institution{Columbia University}
  \streetaddress{116th and Broadway}
  \city{New York}
  \state{NY}
  \country{USA}
  \postcode{10027}
}

\author{Hang Chen}
\authornotemark[1]
\email{hc2798@nyu.edu}
\affiliation{%
  \institution{New York University}
  \streetaddress{6 MetroTech Center}
  \city{New York}
  \state{NY}
  \country{USA}
  \postcode{11201}
}

\author{Doudou Guo}
\authornotemark[1]
\email{dg3195@columbia.edu}
\affiliation{%
  \institution{Columbia University}
  \streetaddress{116th and Broadway}
  \city{New York}
  \state{NY}
  \country{USA}
  \postcode{10027}
}

\author{Yunbai Zhang}
\authornotemark[1]
\email{yz3386@columbia.edu}
\affiliation{%
  \institution{Columbia University}
  \streetaddress{116th and Broadway}
  \city{New York}
  \state{NY}
  \country{USA}
  \postcode{10027}
}

\begin{abstract}


Trading is a highly competitive task that requires a combination of strategy, knowledge, and psychological fortitude. With the recent success of large language models(LLMs), it is appealing to apply the emerging intelligence of LLM agents in this competitive arena and understanding if they can outperform professional traders. In this survey, we provide a comprehensive review of the current research on using LLMs as agents in financial trading. We summarize the common architecture used in the agent, the data inputs, and the performance of LLM trading agents in backtesting as well as the challenges presented in these research. This survey aims to provide insights into the current state of LLM-based financial trading agents and outline future research directions in this field.
\end{abstract}


\received{26 July 2024}

\maketitle




\section{Introduction}

Recent advances in large language models (LLMs) have revolutionized research in natural language processing and demonstrated significant potential in powering autonomous agents \cite{agentsurvey}. LLM agents have been applied across various domains, such as healthcare \cite{agentinclinic} and education \cite{llmedu}. In addition, the finance sector has seen lots of exploration of LLM applications \cite{llminfin, finllmsurvey2}. There has been a emerging trend of developing LLM powered agents for trading in financial markets. Professional traders are required to process amount of information from various sources and quickly make decisions. Therefore, LLMs are well-suited for this role due to their ability to process large amounts of information quickly and produce insightful summaries.

In this survey, we conduct a systematic analysis of the research into using LLMs as agents for financial trading. Our goal is to identify common areas of research and offer insights into future directions. Specifically, we aim to address the following questions:

\begin{itemize}
    \item What are the common architectures in LLM powered trading agents?
    \item What types of data are used for LLMs to make informed trading decisions?
    \item What is the current performance of LLMs in financial trading, along with their potential and limitations?
\end{itemize}

As LLM powered agent is an emerging research topic, relatively few studies have explored applying this technique in financial trading. In this survey, we reviewed 27 papers that study using LLMs for financial trading with seven of which explicitly include the term "agent" in their titles. We identified these papers through multiple Google Scholar search using keywords such as "LLM for trading" and "GPT stock agent." Each paper was manually assessed to confirm its relevance to financial trading with LLM agents. To the best of our knowledge, this is the first paper to review the contributions in the domain of LLM agents for financial trading.

\section{Architecture}
The architecture is a crucial aspect when designing an LLM-based agent, and it's often determined by the agent's objective. Generally, the primary objective of a trading agent is to optimize returns through its trading decisions over a specific period. Besides, other risk-related metrics are also crucial in evaluating agent performance, which we will explore further in Section \ref{sec:eval}. While there are LLM-powered agents designed for various financial tasks such as summarizing financial news \cite{finverse} or acting as financial advisors \cite{llm_advise}, our focus will be on trading agents aiming to achieve investment returns, as this constitutes the majority of research in this area.

\begin{figure*}[h!]
\centering 
\begin{forest}
for tree={          
    draw=blue, thick, rounded corners, minimum height=3ex, minimum width=2em, grow=east, forked edge, 
    s sep=3mm,    
    l sep=4mm,    
    fork sep=3mm  
}
[Finance Trading Agent
    [LLM as an Alpha Miner
        [QuantAgent\cite{quantagent}]
        [AlphaGPT\cite{wang2023alpha}]
    ]
    [LLM as a Trader
        [{RL-Driven:
            SEP\cite{Koa_2024},
            \cite{ding2023integratingstockfeaturesglobal}
        }]
        [{Debate-Driven: 
            TradingGPT\cite{li2023tradinggptmultiagentlayeredmemory}, 
            HAD\cite{xing2024designingheterogeneousllmagents}
        }]
        [{Reflection-Driven: 
            FinAgent\cite{multimodalfinmem},
            FinMem\cite{finmem}
        }]
        [{News-Driven: 
            LLMFactor\cite{wang2024llmfactorextractingprofitablefactors},
            MarketSenseAI\cite{beatunveiling},
            \cite{lopezlira2023chatgptforecaststockprice},
            \cite{llmnewsscore},
            \cite{sentitrade},
            \cite{unveiling}
        }]
    ]
]
\end{forest}
\caption{Overview of architectures of finance LLM agents.}
\label{fig:tree}
\end{figure*}
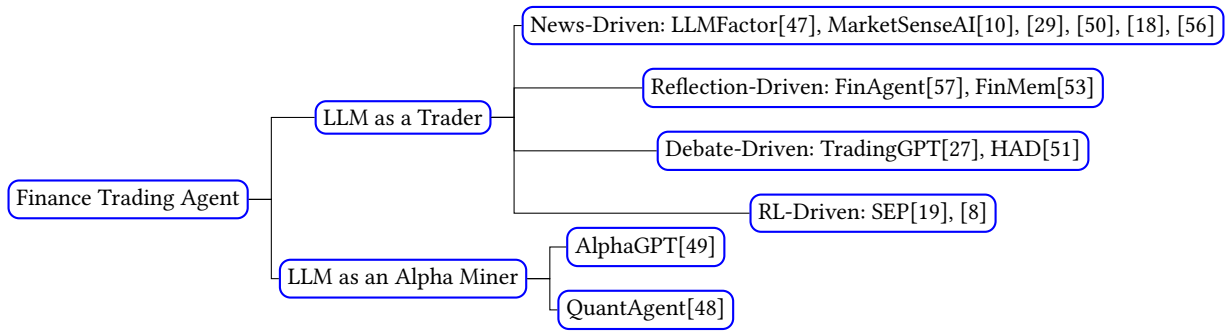

The architectures can be broadly categorized into two types: LLM as a Trader and LLM as an Alpha Miner. LLM trader agents leverage LLMs to directly generate trading decision(i.e. BUY, HOLD, SELL).On the other hand,  Alpha Miner agents utilize LLM as efficient tools to produce high quality alpha factor, which are subsequently integrated into downstream trading systems. A tree diagram demonstrating the hierarchical structure and development of all these architectures is presented in Figure \ref{fig:tree}.

\subsection{LLM as a Trader}
The architecture of LLM trader agents focuses on utilizing LLM to directly make trading decisions. These systems are designed to analyze vast amounts of external data, such as news, financial reports, stock price and refine information from the these data to generate buy or sell signals. This section discusses different sub-types of LLM as a trader agents, including news-driven, reflection-driven, debate-driven and reinforcement learning(RL)-driven agents.
\subsubsection{News-Driven}

News-driven architecture is the most fundamental type, where individual stock news and macroeconomic updates are integrated into the prompt context. The LLMs are then instructed to predict stock price movements for the next trading period. Existing works \cite{lopezlira2023chatgptforecaststockprice, llmnewsscore} evaluate both close-source LLMs(e.g GPT3.5/4) and open-source LLMs(e.g Qwen\cite{bai2023qwentechnicalreport}, Baichuan\cite{yang2023baichuan2openlargescale}, etc.) in financial sentiment analysis. They also backtest simple long-short strategy based on these sentiment scores, demonstrating the effectiveness of trading using such strategy. Additionally, \cite{unveiling,sentitrade} researched the performance of LLMs(FinGPT, OPT, etc.) that specifically fine-tuned with financial related dataset and demonstrated further improvement by aligning LLMs with domain-specific knowledge.

More advanced architectures involve summary, refinement of news data and reasoning of the relationship between news data and stock price movement. \cite{beatunveiling} developed several summary modules, including progressive daily news summary, fundamental and macroeconomic summary, and stock price momentum summary. These summaries are managed by a memorization module and referred as "memory". During the trading stage, relevant "memory" is retrieved as "recommendation" context to generate final trading decision. The authors also found that the general-purpose LLMs such as GPT4\cite{gpt4} has great in-context learning capability in financial oriented tasks. LLMFactor\cite{wang2024llmfactorextractingprofitablefactors} first utilizes LLM's reasoning capability to identify important factors by asking the LLM to analysis relationship between historical news and corresponding stock price movements. Then, the agent extract these factors from daily news and make predictions of stock price during trading.

Note that some approaches are more akin to advanced LLM-based methods, with zero-shot prompting or in-context learning\cite{Li_2023, thepromptreport}, rather than agent-based systems, but they are included here for completeness.

\subsubsection{Reflection-Driven}

Reflection\cite{park2023generativeagentsinteractivesimulacra} is built on extracted memory using LLMs summarization. It is high-level knowledge and insights progressively aggregated from raw memories and observations. Such reflection is used to make trading decisions. In this section, we survey  finance LLM agents that incorporate reflection into their architectures.

FinMem \cite{finmem} introduces a trading agent with layered memorization and characteristics. The raw inputs, such as daily news and financial reports, are summarized into memories. Upon the arrival of new observations, the relevant memories are retrieved and integrated with these observations to produce reflections. Both memories and reflections are stored in a layered memory bucket. During the trading phase, these memories and reflections are retrieved and utilized by the decision-making module to generate the final trading decisions. The retrieval method considers the recency, relevancy, and importance of the information. 

FinAgent\cite{multimodalfinmem} proposed the first multimodal agent with similar layered memory and layered reflection design, with an additional multimodal module that takes in numeric, text and image data. Furthermore, the decision making module incorporates technical indicators such as Moving Average Convergence/Divergence(MACD)\cite{britannica_macd} and Relative Strength Index(RSI)\cite{fidelity_rsi} as well as analyst guidance to effectively capture market dynamics. This framework has demonstrated superior performance in backtesting compared to other agents including FinMem.

The design of memory and reflection can also find its root in cognitive science\cite{chun2011memory}. Analogous to human learning, where human beings interact with the environment, absorb feedback, generate memories  and apply learned lessons to solve tasks, memory and reflection in LLMs based trading agents share similar mechanisms. The inclusion of memory and reflection in LLM-based algorithms offers significant benefits such as  mitigating the risk of hallucinations\cite{ji2023mitigatinghallucinationlargelanguage} and obtaining high-level understanding of the environment\cite{gen_agents_stanford}.

\subsubsection{Debate-Driven}
Debating among LLMs is proven to be an effective method to enhance the reasoning and factual validity. This approach is also widely adopted in LLM financial agent. \cite{xing2024designingheterogeneousllmagents} proposed a heterogeneous debating framework where LLM agents with different roles (i.e. mood agent, rhetoric agent,  dependency agent, etc.) debate with each other, which improves the news sentiment classification performance. TradingGPT\cite{li2023tradinggptmultiagentlayeredmemory} proposed a similar architecture as FinMem\cite{finmem} with one extra step that agents debate on each other's actions and reflections, thereby improving the robustness of reflections.

\subsubsection{Reinforcement Learning Driven}
Reinforcement learning methods, such as RLHF \cite{ouyang2022traininglanguagemodelsfollow} and RLAIF \cite{lee2023rlaifscalingreinforcementlearning}, have proven effective in aligning LLM outputs with expected behaviors. One challenge, however, is obtaining high-quality feedback efficiently and systematically. In financial trading, backtesting provides a cost-effective method for generating high-quality feedback on trading decisions and, intuitively, can serve as a source of rewards in reinforcement learning. SEP \cite{Koa_2024} has proposed leveraging reinforcement learning with memorization and reflection modules in trading agents. This approach utilizes a series of correct and incorrect predictions derived from financial market history to refine the LLM’s predictions in real-world markets.

Furthermore, Reinforcement Learning is well known as a classical method for decision making in games and trading, due to its nature\cite{jiang2017deepreinforcementlearningframework}. \cite{ding2023integratingstockfeaturesglobal} developed a RL-based framework consists of Local-Global (LG) model and Self-Correlated Reinforcement Learning (SCRL), which are made of multi-layer perceptrons. An LLM is used to generate embeddings from news headlines, which are then projected into the stock feature space. These embeddings are integrated with existing stock features to serve as inputs for the LG model. The LG model, functioning as the policy network, is trained via Proximal Policy Optimization (PPO) \cite{ppo} with trajectories sampled from the training trading period.

\subsection{LLM as an Alpha Miner}
Another important category involves agents using LLMs as Alpha Miners, where the LLM generates alpha factors instead of directly making trading decisions. QuantAgent \cite{quantagent} demonstrated this method that leverages the LLM's capability to produce alpha factors through an innerloop-outerloop architecture. In the inner loop, the writer agent takes in a general idea from human trader and generates a script as its implementation. The judge agent provides feedback to refine the script. In the outer loop, the committed code are tested in real world market and the trading results are used to enhance the judge agent. It has been proved that that this approach enables the agent to progressively approximate optimal behavior with reasonable efficiency. In a subsequent research, AlphaGPT\cite{wang2023alpha} propose a human in the loop framework for alpha mining. This approach instantiated an alpha mining agent on a similar architecture and an experimental environment. Both studies demonstrates the effectiveness and efficiency of the LLM powered alpha mining agent system, which is especially valuable as alpha mining is an resource intensive job. 

\begin{figure}[h]
    \centering
    \includegraphics[width=\linewidth]{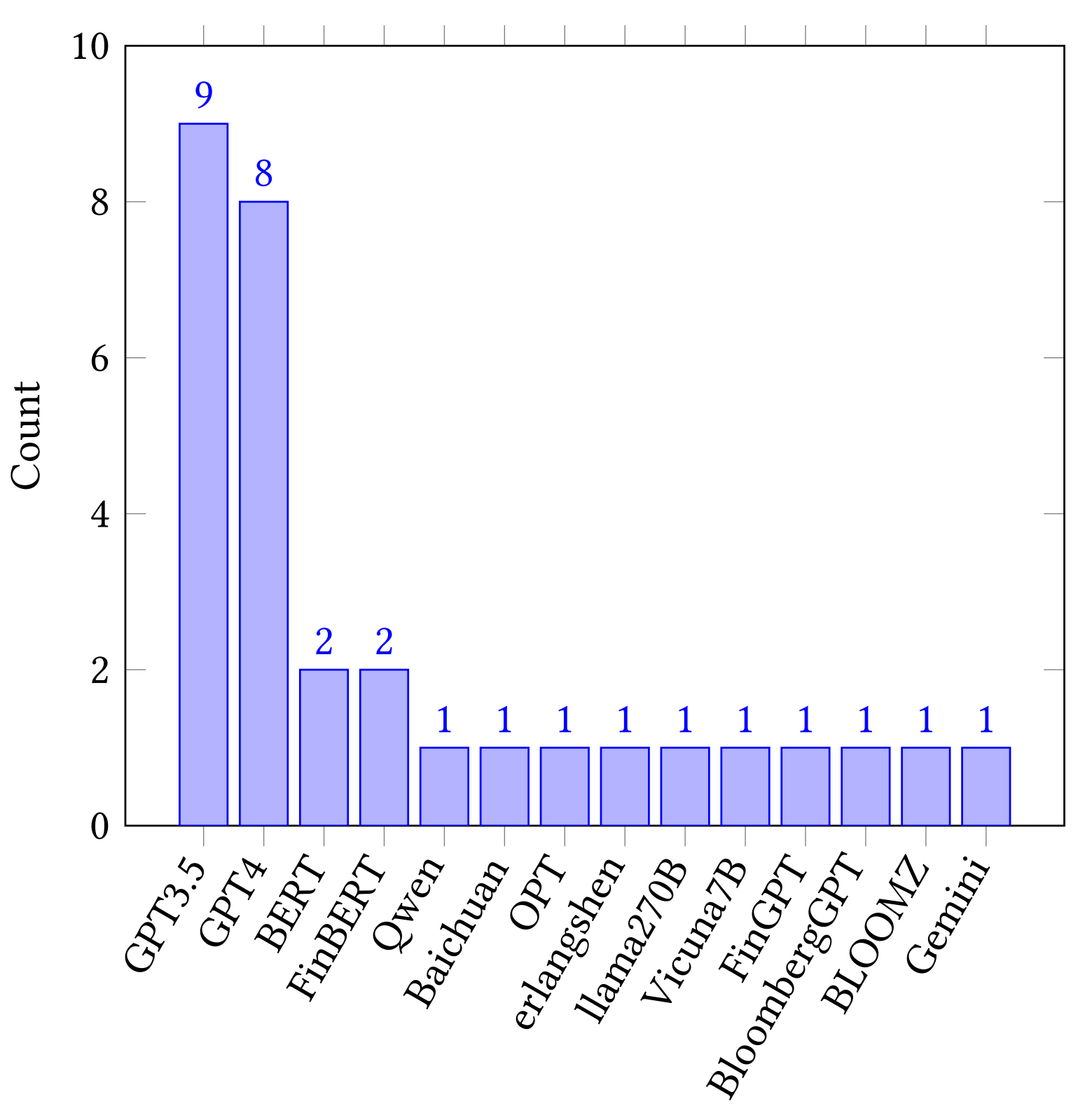} 
    \caption{Histogram of base LLM used by Finance Agent (one paper may contain multiple agent)}
    \label{fig:model_histogram}
\end{figure}

\subsection{LLM Selection in Agent}
Lastly, to investigate the use of different LLM models, we have included a histogram (see Figure \ref{fig:model_histogram}) of the LLM models to provide an overview of the spectrum. It is worth noting that OpenAI's models, particularly GPT-3.5 and GPT-4, dominate research usage due to their outstanding general performance. Also there is a long tail distribution of open source model selection, catering to needs for more flexible and specialized development. Notably, GPT3.5 is even used more frequently than GPT4, indicating a preference on its cost-effectiveness and lower latency. 
%

\section{Dataset}
LLM-powered trading agents rely heavily on diverse data sources to generate trading signals. In our survey, we identified a wide range of data types utilized by various agents, which we have categorized into four major groups:
\begin{itemize}
    \item Numerical Data: Includes numbers or statistics, such as stock prices and trading volumes.
    \item Textual Data: text-based information, such as stock news, financial reports.
    \item Visual Data: Consists of charts and images related to the financial markets.
    \item Simulated Data: Consists of data from simulated stock markets and news events.
\end{itemize}

\subsection{Numerical Data}
In conventional quantitative trading models, numeric data has played a crucial role\cite{quantpaper, stockgpt}. However, LLMs are inherently designed to process textual data. To accommodate numerical data, it must be converted into text strings to ensure compatibility with LLM architectures.  Despite LLMs' known weaknesses in arithmetic problem and reasoning\cite{hendrycks2021measuringmathematicalproblemsolving}, numerous studies have successfully incorporated numerical data into into LLM-based trading agents. \cite{multimodalfinmem} calculates common stock price features, such as three-day price changes, from raw stock price data. These features are then described and summarized by the LLM to form short-term, mid-term, and long-term signals. These signals contribute to the low-level reflection processes of the LLM agent.

In \cite{quantagent}, the authors further used additional numerical market data such as open and close price, high and low price to create trading ideas that guide the outer feedback loop. This data also serves as a means of evaluating the generated alpha strategy. Our findings suggest that incorporating numerical data is essential, as it inherently reflects the characteristics of the financial market. For example, high trading volumes and rising prices typically signify positive market expectations, often correlating with company performance.

%

\subsection{Textual Data}
Textual data such as news or financial reports are critical to financial traders. We found all LLM powered agents reviewed in this paper use textual financial data as input. Based on the terminology commonly used in the financial industry, we categorize textual data into two types: Fundamental Data and Alternative Data.

\subsubsection{Fundamental Data} 
Fundamental data encompasses information that represents the primary characteristics and financial metrics for assessing the stability and health of an asset. Fundamental data used in LLM trading agents includes financial reports and analyst reports.

\paragraph{Financial Reports}
Financial reports, such as Form 10-Q and Form 10-K filings, are critical for understanding a company's performance. These documents provide LLM agents with insights into corporate financial status, performance, and future expectations. They are extensively utilized by financial trading agents like FinMem \cite{finmem}, TradingGPT \cite{li2023tradinggptmultiagentlayeredmemory}, and FinAgent \cite{multimodalfinmem}. These works incorporate financial reports to enrich the agents’ memory and make informed trading decisions.

\paragraph{Analyst Reports}
In addition to financial statements, analyst reports and investment research from industry professionals provide invaluable data. These sources offer high-quality insights, opinions, and forecasts beyond the information found in public financial reports and news articles. For example, FinAgent \cite{multimodalfinmem} incorporates expert guidance from SeekingAlpha\footnote{https://seekingalpha.com/} as a crucial input to its decision-making module. It is used with other data sources such as market intelligence, analysis of price movements, and historical trading decisions.

\subsubsection{Alternative Data}
Alternative data refers to non-traditional information used to evaluate companies and markets. This type of data complements traditional sources such as financial reports. By leveraging alternative data, trading agents can obtain unique perspectives on various issues, thereby enhancing their investment decision-making processes.

\paragraph{News Data} 
News data from reputable sources such as Bloomberg\footnote{https://www.bloomberg.com/}, The Wall Street Journal\footnote{https://www.wsj.com/}, CNBC Television\footnote{https://www.cnbc.com/}, or stock research platforms provides real-time information on market movements, industry trends, and company-specific developments. This type of data is extensively used in various studies \cite{li2023tradinggptmultiagentlayeredmemory, multimodalfinmem, finllama, finmem, xing2024designingheterogeneousllmagents} to stay up-to-date with the real-world financial market. Specifically, LLMs excel at extracting sentiment information from news data, which could a crucial signal for trading decisions.

\paragraph{Social Media Data} 
In addition to traditional news sources, researchers can also use of social media data, such as Twitter, StackExchange, StockTwits and Reddit posts, to capture more informal, real-time discussions about financial topics. There are many machine learning models that ultilize social media data for stock price prediction such as \cite{twitter1, reddit, socialmediamodel}. However, SEP \cite{Koa_2024} is the only work we've reviewed that incorporates real-time social media data in LLM trading agent. In SEP, LLM is used to generate and summarize key facts from twitter data of a given stock. Incorporating social media data is a under studied field but with great potential.

\subsection{Visual Data}

Numerical and textual data have been predominately used in trading agent design, while visual data has been less explored as an additional data source. One of the reason for this disparity stems from the challenges that existing LLM models have in effectively processing and understanding financial visual data. While recently proposed multimodal LLMs such as LLaVA \cite{liu2023visualinstructiontuning}, GPT-4v\cite{2023GPT4VisionSC} possesses the capability to processing visual data, most of these models have not been specifically trained and evaluated on financial visual data such as Kline charts, volume charts. An early experiment by FinAgent \cite{multimodalfinmem} in incorporating visual data using GPT-4v in the trading agent context has shown promise. FinAgent integrates Kline Charts and Trading Charts along with numerical and textual data. It demonstrated significant trading performance improvements over FinMem which uses a similar architecture but without visual input. This effort represents a significant step forward in utilizing visual data within LLM frameworks for trading applications, by leveraging trading chart information, which forms the cornerstone of technical analysis widely used by traders. This pioneering work sets a promising direction for further exploration in integrating visual data into LLM-based trading agents.

\subsection{Simulated Data}
Simulated data and environments are created to replicate real-world scenarios, providing finance professionals with effective tools for understanding both market dynamics and LLM agent behavior. In \cite{aimeetsfinance}, a group of LLM stock agents with varying personalities engage in trading within a simulated environment. This simulation includes not only market price fluctuations but also synthetic events such as interest rate changes and the release of financial reports. Additionally, agents can communicate via a Bulletin Board System. Experiments have shown that agents with different personalities react differently, and external factors significantly influence their behavior.

Simulated data is also invaluable for researching LLM behaviors concerning bias, ethics, and robustness under extreme circumstances in a controlled manner. \cite{scheurer2024large} defined several real-world scenarios including one with extreme pressure to examine LLM agents' behavior in these circumstances. The study revealed that LLMs are capable of taking unethical actions under high-pressure conditions, such as using insider information to trade for profit and even crafting deceptive explanations to conceal such actions. This study underscores the potential regulatory risks associated with using LLMs in financial trading. Therefore, it is imperative to thoroughly investigate such issues before deploying them in a production environment.

\section{Evaluation}
\label{sec:eval}
In the papers we have surveyed, LLM powered trading agent have demonstrated superior performance during backtesting. In this section, we discuss trading strategies generated by LLM agents, as well as the evaluation metrics and baselines used to assess the performance of LLMs via backtesting.

\subsection{Trading Strategy}

LLM generates simple trading signals such as "Buy", "Hold", "Sell" by analyzing textual data like market news or financial statements. In FinMem\cite{finmem} and FinAgent\cite{multimodalfinmem}, the signal is directly used for trading action for a particular stock. However, when managing a portfolio with multiple stocks, a common approach is to use ranking-based strategies. These strategies require a numeric score to rank the stocks and allocate funds based on the magnitude of these scores. In FinLlama\cite{finllama}, all stocks in S\&P500 index are ranked by an LLM and top 35\% are assigned to long position while the bottom 35\% are assigned to the short position. A similar approach is adopted in \cite{sentitrade, lopezlira2023chatgptforecaststockprice}, where the long-short strategy has shown to outperform both long-only and short-only strategies in backtesting. On the other hand, \cite{llmnewsscore} allocates long positions to stocks with overall positive news sentiment and short positions to those with negative sentiment, without considering the magnitude of the sentiment scores. This approach does not fully utilize the signal, leading to the observation that the long-short strategy performed worse than the long-only strategy in their experiment. In \cite{unveiling}, stocks are grouped based on their signal rankings, with the top-ranked group showing the best returns compared to others. 

In executing trading strategies, stocks are typically weighted either equally or based on market capitalization size. In both\cite{sentitrade} and \cite{lopezlira2023chatgptforecaststockprice}, portfolios weighted by market capitalization have shown slightly higher returns than those that are equally weighted. We conjecture that the quality of textual signals from large-cap companies is better than that from smaller companies due to the bias in news coverage.

\subsection{Metrics}
\paragraph{Portfolio Performance Metrics} Almost all works we surveyed uses common performance metrics in evaluating the trading agent. Cumulative return and annualized return are used to measure overall profitability of the trading strategies. Sharpe Ratio \cite{sharpe1966} and Maximum Drawdown are used to assess the risk of the trading performance. One observation we had is that while both risk and profit metrics are commonly used, few studies consider trading costs in their evaluations.

\begin{itemize}
    \item Cumulative Return:\[
\text{Cumulative Return} = \left( \frac{P_t - P_0}{P_0} \right) \times 100\%
\]
where:
\begin{itemize}
  \item $P_t$ is the ending price (or value) at time $t$
  \item $P_0$ is the initial price (or value) at the beginning
\end{itemize}
    \item Annualized Return
\[
\text{Annualized Return} = \left( \frac{P_t}{P_0} \right)^{\frac{1}{t}} - 1
\]
where:
\begin{itemize}
  \item $P_t$ is the ending price (or value) at time $t$
  \item $P_0$ is the initial price (or value) at the beginning
  \item $t$ is the number of years
\end{itemize}

\item  Sharpe Ratio:
\[
\text{Sharpe Ratio} = \frac{R_p - R_f}{\sigma_p}
\]
where:
\begin{itemize}
  \item $R_p$ is the return of the portfolio
  \item $R_f$ is the risk-free rate
  \item $\sigma_p$ is the standard deviation of the portfolio's excess return
\end{itemize}

\item Maximum Drawdown up to time $T$ is given by:
\[
\text{MDD}(T) = \max_{\tau \in (0,T)} D(\tau) = \max_{\tau \in (0,T)} \left[ \max_{t \in (0,\tau)} X(t) - X(\tau) \right]
\]
where:
\begin{itemize}
  \item $X(t)$ is the value of the portfolio at time $t$
  \item $X(\tau)$ is the value of the portfolio at time $\tau$
  \item $T$ is the time period being considered
\end{itemize}
    
\end{itemize}

\paragraph{Signal Metrics}
Sometimes, portfolio performance metrics do not directly reflect the performance of a trading agent or the effectiveness of a trading signal. Therefore, it is equally important to monitor the predictive power of the generated signals. In \cite{sentitrade} and \cite{usechatgptbetter}, the F1 score and accuracy are used to measure the model's prediction accuracy of news sentiment. Meanwhile, \cite{beatunveiling} and \cite{unveiling} use the win rate to measure the proportion of profitable trades out of all executed trades. In QuantAgent \cite{quantagent}, the Information Coefficient (IC) \cite{ic} is calculated to quantify the correlation between predicted signals and future returns.

\paragraph{System Metrics}
Utilizing LLM-powered trading agents to process information and generate trading signals often involves leveraging commercial LLM APIs such as ChatGPT\footnote{https://chatgpt.com/}. However, QuantAgent is the only study we have encountered that addresses the cost of generating LLM tokens and the computational time complexity for both training and inference. This could be due to the fact that the cost of token generation is usually negligible compared with the capital size of the portfolio.

\subsection{Backtest Setting}

\begin{table}[htbp]
  \centering
  \begin{tabular}{p{4cm} p{3cm}}
    \toprule
    \textbf{Number of Years} & \textbf{Count of Papers} \\
    \midrule
    0$\sim$2 & 8 \\
    2$\sim$5 & 2 \\
    $\geq$5 & 4 \\
    \bottomrule
  \end{tabular}
  \caption{Number of Years Covered in Backtest Testing}
  \label{tab:backtest}
\end{table}

To evaluate the performance of LLM powered agents, most of the work use backtesting with real market data. For agents evaluated on single-stock portfolios, stocks with the highest volume of accessible news data are seleted for testing. For example, stocks such as TSLA, AMZN, MSFT, COIN, NFLX, GOOGL, META, PYPL selected to trade in \cite{finmem}, \cite{multimodalfinmem} and \cite{usechatgptbetter}. For agent managing multi-stock portfolios, index component stocks are typically selected, such as those from the SP500\cite{sp500} and CSI300\cite{csi300}.

Most agent-based models are backtested exclusively on the stock market. Among the 14 papers that use real market data for backtesting, 9 focus on the US stock market and 5 on the Chinese market. Only FinAgent \cite{multimodalfinmem} extends its backtesting on the cryptocurrency market, specifically trading ETH \cite{ethereum}.

We also observe that most evaluations set the backtesting period between 2020 and 2024, coinciding with the publication date of the work. On average, the median of testing period is only 1.3 years (Table \ref{tab:backtest}), with the exact start and end dates chosen rather arbitrarily. While LLM agents have demonstrated strong performance during backtesting, a short and single backtesting period may diminish the credibility of the results.

\subsection{Baseline and Performance}
During backtesting, the baselines methods can be divided into 3 major categories: Rule based, Machine Learning (or Deep Learning) based and Reinforcement Learning based. In \cite{finmem, multimodalfinmem, stockgpt}, rule based strategy such as "Buy and Hold", "Mean Reversion" and "Short-Term Reversal" are used as baseline. Given that classification models can be used for news sentiment prediction, machine learning or deep learning models such as Random Forest\cite{randomforest}, LightGBM\cite{NIPS2017_6449f44a}, LSTM\cite{LSTM}, and BERT\cite{bert} are also used as baselines. Furthermore, Reinforcement Learning algorithm are increasingly popular in quantitative trading \cite{li2023generalframeworkenhancingportfolio}. Deep reinforcement learning frameworks such as PPO\cite{ppo} and DQN\cite{dqn} are also used as a benchmark in \cite{finmem, multimodalfinmem}. 

Overall, LLM powered trading agents have demonstrated strong performance in backtesting. Our survey\cite{finmem, multimodalfinmem, ding2023integratingstockfeaturesglobal, lopezlira2023chatgptforecaststockprice} shows that LLM agents have achieved annualized return ranging from 15\% to 30\%  over the strongest baseline during backtesting period with real market data, which demonstrates the great potential of using LLM in financial trading.

\section{Limitation and Future Direction}
Although the use of LLM agents in financial trading has achieved many successes, limitations still exist in current research.

From an architectural perspective, most agents rely on closed-source models (e.g., GPT-3.5/GPT-4), which raises concerns about data privacy and restricts the ability to customize model development. Additionally, our review reveals that most studies applies LLMs through in-context learning without any fine-tuning, with only \cite{Koa_2024} tunes the LLM during training. The effectiveness of fine-tuning LLMs for trading agents remains an open question. Another significant issue is the inference latency, which can be a bottleneck, making these models impractical for high-frequency trading. Moreover, integration with existing trading systems is rarely discussed in the literature we surveyed.

From a data perspective, while agents typically use textual data such as news and fundamental data, few utilize social media data, which can significantly influence financial market (i.e. the Game Stop Short Squeeze \cite{gamestop}). 

From an evaluation perspective, backtesting is predominantly confined to the US and Chinese stock markets, with notable absences in other financial markets such as derivatives, bonds, or commodities. Additionally, backtesting periods are generally short, and few studies consider trading costs. Expanding evaluations to include these other markets and accounting for trading costs could unveil new opportunities, particularly given the potential sensitivity to news data in these markets.

Lastly, agents with different trading style or personality tends to perform differently in making trading decisions. However, few studies have conducted ablation studies to explore the underlying reasoning processes of LLMs in their trading decisions. Utilizing simulated environments could be a promising approach to gain deeper insights into the LLMs' decision-making processes and patterns.

\section{Conclusion}

In this survey, we systematically reviewed all relevant works that leverage LLMs as trading agents, focusing on their architectural design, data inputs, and evaluation methods. Although this is an emerging field with relatively few studies to date, we found that LLM-powered trading agents demonstrate significant potential in extracting signals from massive amount of textual information and making informed decisions. However, there are still challenges including the reliance on closed-source models and the integration issues with existing trading systems and human traders, which can be important directions for future research.

\bibliographystyle{ACM-Reference-Format}
\bibliography{sample-base}


\end{document}